\documentclass{ws-procs975x65}

\begin{document}

\title{Dynamics of metal clusters in rare gas clusters}

\author{M. BAER}

\address{Institut f{\"u}r Theoretische Physik, Universit{\"a}t Erlangen,
Staudtstrasse 7, \\ D-91058 Erlangen, Germany}

\author{G. BOUSQUET, P.M. DINH$^{*}$}

\address{Laboratoire de Physique Th\'eorique, UMR 5152, Universit\'e
  P. Sabatier 
118 Rte de Narbonne,\\ F-31062 Toulouse cedex, France\\
$^{*}$E-mail: dinh@irsamc.ups-tlse.fr}

\author{F. FEHRER, P.-G. REINHARD$^{**}$}
\address{Institut f{\"u}r Theoretische Physik, Universit{\"a}t Erlangen,
Staudtstrasse 7, \\ D-91058 Erlangen, Germany ,\\
$^{**}$E-mail: mpt218@theorie2.physik.uni-erlangen.de}

\author{E. SURAUD}
\address{Laboratoire de Physique Th\'eorique, Universit\'e P. Sabatier
118 Rte de Narbonne,\\ F-31062 Toulouse cedex, France}

\begin{abstract}
We investigate the dynamics of  Na clusters embedded in Ar matrices. 
We use a hierarchical approach, accounting microscopically 
for the cluster's degrees of freedom and more coarsely for the matrix. The
dynamical polarizability  of the Ar atoms and the strong Pauli-repulsion 
exerted by the Ar-electrons are taken into account.
We discuss the impact of the matrix on the cluster gross properties and 
on its optical response. We then consider a realistic case of irradiation
by a moderately intense laser and discuss the impact  of the matrix 
on the hindrance of the  explosion, as well as a possible pump probe scenario 
for analyzing dynamical responses. 
\end{abstract}

\keywords{Metal cluster, rare gas matrix, 
laser irradiation, Density Functional Theory}

\bodymatter

\section{Introduction}\label{aba:intro}

Structural and dynamical properties of clusters,
especially in the case of  simple alkali clusters,
have focused many experimental as well as theoretical investigations
\cite{Kre93,Hab94a,Hab94b,Eka99,Rei03a}. The case of clusters 
embedded in a matrix or deposited on a surface requires 
specific treatments, because the environment
influences cluster properties and has thus to be incorporated in the
analysis.  It should nevertheless be noted that the 
presence of an environment (surface or matrix) may
simplify the experimental handling: it usually implies
localization, fixed geometry,  well controlled temperature, and higher
yields. Deposited and/or
embedded clusters thus often represent a preferred alternative for cluster
studies. But the presence of an
environment complicates the theoretical modeling. It requires 
a handling of the cluster-matrix interface and the huge number of
atoms in the matrix or substrate may become prohibitive. 
But because  many interesting experiments
have been, or can only be, done with clusters in contact with a
carrier material such theoretical investigations, 
although challenging, become unavoidable.  
Let us mention as examples of experimental motivations the systematics of
optical response in large noble-metal clusters \cite{Nil00,Gau01} and
its dependence on the environment \cite{Die02}. 

While static properties of embedded/deposited clusters have already been
studied since long \cite{Hab94a,Hab94b}, 
dynamical scenarios still remain little explored \cite{Rei03a} but promise
challenging questions. As in free clusters, the optical
response represents a key tool of analysis, both of cluster structure, in
particular in  metal clusters \cite{Kre93} and 
in most dynamical situations.  Let us mention the case of
moderate perturbations, as, {\it e.g.}, in the
case of laser irradiation in which resonant coupling between laser
and cluster eigenfrequencies represents a basic doorway mechanism for
many dynamical scenarios \cite{Rei03a}. The
study of the optical response thus constitutes a key "entry point" for
the cluster response to electromagnetic probes.

Among the various possible combinations of clusters with
substrate/matrix, the case of inert or moderately active environments
is especially interesting. It implies only moderate
perturbations of cluster properties and one can thus still benefit from
the well defined geometry of the system and  access
predominantly the cluster properties themselves. From the theoretical
side, the environment can be included at a
lower level of description (hierarchical method), 
which simplifies the handling. This is the type of situations 
we shall focus on in the following, namely the  case of simple metal clusters
(sodium) inside a moderately active environment (rare gas matrix). 

The term "hierarchical method", used above, requires some comments.
It refers to the fact that the interface may be treated at a simpler
level than the cluster's degrees of freedom,
because of the moderate interactions between
cluster and environment. Such approaches are
well known in many complex systems, as for example the shell model in surface
chemistry \cite{Mit93a,Nas01a} or the coupled quantum-mechanical with
molecular-mechanical method (QM/MM) often used in
bio-chemistry \cite{Fie90a,Gao96a,Gre96a}. 
Hierarchical modeling is nevertheless not free from difficulties, especially 
at the technical side, because the
interface interaction requires a very careful adjustment.
And this may become very tough when one is dealing with a
huge range of energies (cluster {\it vs.} matrix), as in the case of 
an alkaline cluster embedded in a rare gas (RG) matrix, where
one has to deal with eV (Na) as well as meV (RG-RG and RG-Na).
But once properly tuned  the model becomes able to simulate 
rather easily realistic
situations, especially in terms of system's sizes.

The paper is organized as follows. A brief presentation of the
hierarchical approach is given in section \ref{sec:model}. 
We then discuss the impact of the matrix on gross properties 
of a small metal cluster. We then pursue the analysis by considering 
the impact of the matrix on the optical response of the embedded cluster.
We finally illustrate possible violent irradiation scenarios on a
realistic example. 

\section{Model}
\label{sec:model}

The model has been 
presented in detail in  \cite{Feh05a} 
and we just remind here the basic ingredients. 
The Na cluster is described in the TDLDA-MD approach, 
an approach which has been
validated for linear and non-linear dynamics of free metal clusters
\cite{Rei03a,Cal00}.   
It combines a  (time-dependent) density-functional theory at the 
level of the local-density
approximation (TDLDA) for the electrons, 
 to a classical molecular (MD) dynamics for the ions. The electron-ion interaction
is described by soft, local pseudo-potentials \cite{Kue99}. 
Two classical degrees-of-freedom are associated with each Ar atom : its 
center-of-mass and its electrical dipole moment. The dipoles allow to 
explicitely treat the dynamical polarizability of the atoms
with help of  polarization potentials \cite{Dic58}. The
atom-atom interactions are described by a standard Lennard-Jones potential,
and for the Ar-Na$^+$ subsystem we employ effective potentials from
the literature \cite{Rez95}.
The pseudo-potential for the electron-Ar core repulsion has been
modeled in the form proposed by \cite{Dup96}, and we 
slightly readjust it  by a final
fine-tuning to the NaAr molecule as benchmark (bond length, binding
energy, and optical excitation).
A Van der Waals interaction is also added and computed {\it via} the variance
of dipole operators \cite{Ger04b,Feh05a,Dup96}.

The starting quantity is the total energy which is constructed from the various pieces as outlined 
above. From the total energy, the corresponding equations of motion
can be derived by standard variation. This leads to the time-dependent
Kohn-Sham equations for the cluster electrons. And one obtains 
Hamiltonian equations of motion for the classical degrees 
of freedom (namely Na$^+$ ions and Ar
atom positions and dipoles). The initial condition
is  obtained  from the corresponding stationary equations. 
The Ar-sodium configuration is produced as follows. One cuts a
finite piece from  a fcc Ar crystal and optimizes the structure. 
One then carves a central
cavity to place the Na cluster inside.
The total configuration is then re-optimized  by means of cooled
molecular dynamics for the ions and atoms coupled to stationary
Kohn-Sham solution for the cluster electrons. 

We compute several observables to analyze the statics 
and dynamics of the system.
The ionic and electronic cluster structure is characterized 
in terms of r.m.s. radii
\begin{equation}
r_{I,e}= {\sqrt{<{(x^2+y^2+z^2)}>_{I,e}}}
\label{eq:rms}
\end{equation}
where $<.>_I= \sum_I ./N_I$ and $<.>_e=\int d^3r .\rho_e({\bf r}) /N_e$.
We also analyze the shape of the system by computing the 
reduced quadrupole moment $\beta_2$ which reads
\begin{equation}
\beta_2= \frac{4\pi}{5} \frac{\sqrt{\frac{5}{16\pi}} <(2
z_I^2-x_I^2-y_I^2)>_{I,e}} {<r^2>_{I,e}}
\label{eq:b2}
\end{equation}
with obvious notations.
The optical response is obtained from the dipole momemt $D_{\rm
el}=\langle\sum_n z\rangle$ as quantum mechanical expectation value over the
Kohn-Sham state. The dynamics is 
initiated by an instantaneous dipole boost of the cluster
electrons. 
The dipole signal is followed in the time domain $D_{\rm el}(t)$ and then 
Fourier transformed to the frequency domain  to
obtain the dipole strength as imaginary part of the Fourier transformed dipole
moment \cite{Cal95a,Cal97b,Yab96}.

The numerical solution proceeds with standard methods as 
described in detail in \cite{Cal00}.
The (time-dependent) Kohn-Sham equations for the cluster electrons are
solved using real space grid techniques. The time 
propagation proceeds using a time-splitting method, and  the stationary
solution is attained by accelerated gradient iterations. 
We furthermore employ the cylindrically-averaged pseudo-potential scheme
(CAPS) as introduced in \cite{Mon94a,Mon95a}, which is 
justified for the chosen test cases. The Na$^+$ ions
as well as the Ar atoms are nevertheless still treated in full 3D.

\section{Gross properties of embedded clusters}

\begin{figure}[htbp]
\begin{center}
\psfig{figure=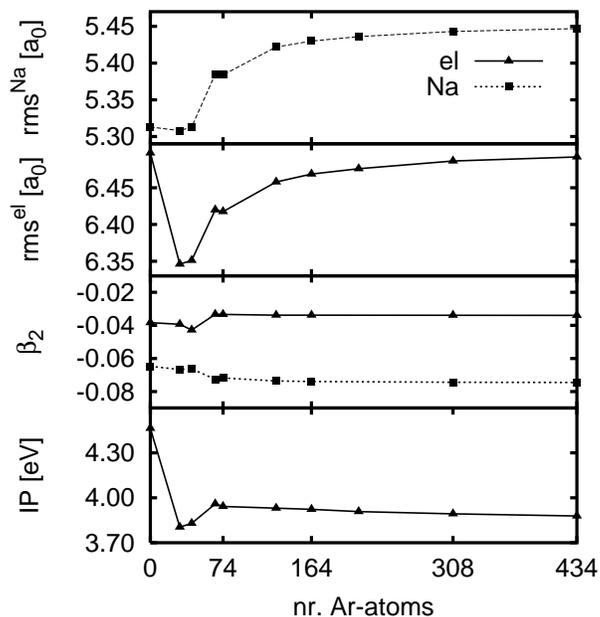,height=82mm,angle=-90}
\end{center}
\caption{
Trend of basic observables with matrix size:
ionic and electronic r.m.s. radius, ionic and electronic quadrupole
deformation $\beta_2$, electronic ionization potential (IP).
The matrix is optimized anew for each configuration.
From \cite{Feh05a}.}
\label{fig:Na8_basobs}
\end{figure}
We take as test case a Na$_8$ cluster embedded into an Ar matrix of
various sizes. Fig. \ref{fig:Na8_basobs} summarizes a few basic
properties of the embedded cluster as a function of increasing  matrix
size. For moderate size matrices the electronic radius is first
reduced (as compared to the free cluster), because of the strong
repulsive Ar-core potentials. But with increasing matrix size both the
electronic and the ionic radii expand in a similar way because of the
monopole-dipole interaction between ${\rm Na}^{+}$ and the
Ar-atoms. The global deformation (third panel from top in Fig.
\ref{fig:Na8_basobs}) is nearly independent of matrix size, although
one can spot a slight trend for small sizes.  The ionization
potential (IP) is, as usual,  defined as the energy difference between
Na$_8$ and Na$_8^+$ for ionic and atomic positions frozen. The IP
globally exhibits a decrease mostly due to the short range
compression. There is indeed a big jump from the free to the clusters
inside small matrices and very little changes amongst all embedded
systems, in accordance with the short-range nature of the Ar core
potentials.  Generally speaking one can identify  from the figure  two
regimes, with a threshold at 164  Ar atoms. Below that limit one
observes specific size effects, which tend to be smoothened  out for
larger sizes for which one observes a monotonous trend towards the
asymptotic value.  This again reflects the balance between the two
competing effects, namely the short range  repulsion due to
compression effects and the long range attraction due to
polarization effects.



\section{Linear optical response}

\begin{figure}[htbp]
\begin{center}
\psfig{figure=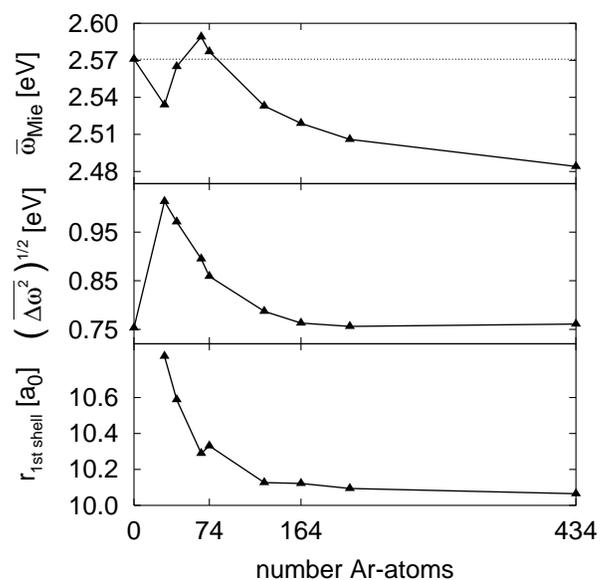,height=82mm,angle=-90}
\end{center}
\caption{
Average peak position, width and radius of the innermost Ar shell, as
a function of matrix size. From~\cite{Feh05a}.
}
\label{fig:Na8_optresp}
\end{figure}

Fig.~\ref{fig:Na8_optresp} presents the trend of the average plasmon
peak position $\bar{\omega}_{\rm Mie}$ and width $(\overline{\Delta
  \omega^2})^{1/2}$ with matrix size, as well as the radius of the
innermost Ar shell. The value $\bar{\omega}_{\rm
  Mie}$ is the averaged dipole transition over the interval $[1.9 \;
  {\rm eV};3.2 \; {\rm eV}]$. The width is the variance of the frequency
computed in the same averaging interval.
Once again, visible structures are seen only below 164 Ar atoms. While
the ${\rm Na}_8$ exhibits a clean Mie plasmon peak, spectral structures of
the clusters embedded in small matrices 
 are strongly fragmented, corresponding to a larger
width (see middle panel). This effect is clearly due to Landau
fragmentation, since the  
surrounding Ar atoms modify the spectral density of
one-electron-one-hole states near the resonance. Indeed the latter
emerges from a subtle interplay of core repulsion and dipole attraction. 
For matrices larger  than 164 Ar atoms, the variance decreases rapidly
and saturates at the same value as that of the free ${\rm Na}_8$,
leading to a fairly clean plasmon peak again.

Note however that the changes in the average
resonance position are very little or even insignificant, that is,
less than 
1/10 of an eV. This is the result of an almost complete cancellation
of core and polarization potentials \cite{Feh05a}.
Indeed we have computed the dipole
spectrum for Na$_8$\@Ar$_{164}$ while switching off the polarization
potentials from the Ar atoms. Thus the cluster electrons feel only the
repulsion of the Ar core potentials. The effect is a strong blue-shift of the
resonance by about 0.8 eV. Switching on the polarization potentials produces
an equally strong red-shift. At the end, the peak position comes out almost
unchanged as compared to the free cluster. It is obvious that the final
resonance position, resulting from two large and counteracting effects,
is extremely sensitive to details of the model. One should thus be careful not to
overinterpret the trends at the meV scale. 

Looking back to Fig.~\ref{fig:Na8_optresp} with that mechanism in
mind, the radius of the innermost Ar shell shrinks strongly up to
about 74 atoms (bottom panel). This corresponds to the regime where core
repulsion increases. The dipole attraction increases as well and we
see practically constant average peak position.  For 
larger systems, whereas the core effect stabilizes, the dipole
attraction still increases, although on a lower rate. This explains why 
a small but steady red-shift is seen
for further increasing system sizes. We have estimated the effects for
further shells towards the full crystal and we found about 14 meV more
red-shift asymptotically. It is a
systematic effect, but a very small one. Thus we will ignore it for
the now following explorations of violent excitations.

\section{Strongly excited Na cluster in an Ar environment}

In this section, we present a realistic intense laser
excitation of the system 
${\rm Na}_8@{\rm Ar}_{434}$. The laser features are the following:
polarization along the symmetry axis of the system (and denoted below
as the $z$ axis),
frequency of 1.9 eV, intensity of $2\times 10^{12}$ W.cm$^{-2}$ and
FWHM of 50 fs. The system reacts mainly by a strong
dipole moment which leads to a direct emission of 3 electrons which
escape instantaneously (i.e. within 3-10 fs). One may wonder whether 
the strong external or effective Coulomb field
at Ar sites would also trigger electron emission from Ar. This was
checked by TDLDA calculations for an Ar atom and we found a critical
field strength of about 0.1 Ry/a$_0$ for that process. In our
calculations, we record the actual field strengths at all Ar
sites. They stay safely below this critical value all times.  Electron
emission thus comes exclusively from the Na cluster (Na$_8$ $\rightarrow$
Na$_8^{3+}$). 

\begin{figure}[htbp]
\begin{center}
\psfig{figure=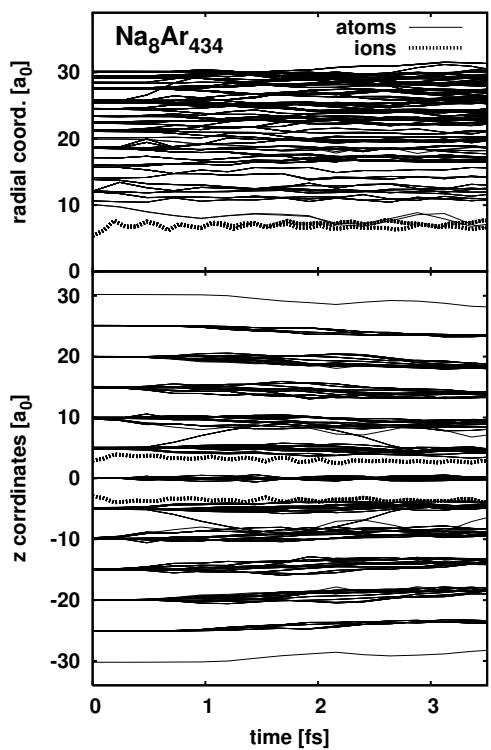,height=100mm}
\end{center}
\caption{
Time evolution of the the atomic (full lines) and ionic (dotted lines)
$z$-coordinates (lower panel) and radial distances
$r=\sqrt{x^2+y^2+z^2}$ (upper panel) for Na$_8$\@Ar$_{434}$ excited
with a laser (see text for details).}
\label{fig:coulomb}
\end{figure}

In Fig.~\ref{fig:coulomb}, showing the time evolution of the Na ions
and Ar atoms coordinates, we observe a rearrangement of the whole
system at ionic/atomic time scale, due to 
the thus produced large Coulomb pressure.
We note however different interesting time scales. First the Na ions
start, up to about 200 fs, a Coulomb explosion, almost
identical to the similar case of a free Na$_8$ cluster lifted to
charge 3$^+$. Second when the ions hit
the repulsive core of the first shell of Ar, 
the ``explosion'' is abruptly stopped. Then the ionic motion
turns to damped oscillations around a (r.m.s.) radius of about 7
a$_0$. Note that the shape
oscillations of the cluster have the typical cycle of about 250 fs,
well known for free Na clusters~\cite{Rei02b} and re-established for
embedded ones~\cite{Feh05b}.

Two stages in the matrix are also seen, although on longer time
scales. We first observe a ``diffusion'' of the
perturbation (due to rearrangement of Na ions) into the various
Ar shells. This is especially visible along the $z$ axis which allows
to read off the propagation speed of this perturbation as 20-30
a$_0$/ps.
We have estimated the sound velocity in the corresponding large pure
Ar cluster (Ar$_{447}$) by computing its vibrational spectrum. The
radial compression mode corresponds to the sound mode in bulk
material. We found a frequency of $\omega_{\rm vib}=1.8$ meV.  The
momentum 
of the radial wave is $q=\pi/R$ where $R=30$ a$_0$ is the cluster
radius. The sound velocity is then estimated as $v_{\rm
sound}=\omega_{\rm vib}/q\approx30$ a$_0$/ps which is very close to
the propagation speed as observed in the figure. One may thus 
interpretate the perturbation as a sound wave sent by the
initial bounce of the Na ions.

Then the perturbation generates oscillations combined with some
diffusion which, after about 1.5 ps, has spread over all shells.
Even the outermost shell is perturbed and the matrix seems to get an
oblate deformation, as the Na$_8$ does. However whereas the Na cluster
exhibits a 
global relaxed trend to oblate shapes, the relaxation of the
Ar atoms seems much longer than that of the Na ions and far beyong the
time scale computed here. These long time
scales for full relaxation and evaporation of Ar atoms are well known
from experiments of dimer molecules embedded in Ar clusters, see
{\it e.g.} \cite{Vor96a}.

Let us now briefly comment on the finite net charge of our system. It 
emerges because the electrons propagate almost unhindered through the
surrounding Ar cluster and eventually escape to infinity. We may
wonder whether, in a macroscopically large matrix, the electrons would
be stuck somewhere 
in the range of their mean free path, drift very slowly back towards
the now attractive cluster well, and eventually recombine there. This
has been checked by using reflecting boundary conditions rather
than absorbing ones. Up to 4 ps, no recombination has been observed.
Furthermore the ionic/atomic rearrangements following
the irradiation turn out to be qualitatively very similar whatever the
boundary conditions. Thus the present scenario should provide a
pertinent picture for a few ps, the time window studied here.

Finally one may wonder how to analyze the key pattern of the
embedded cluster dynamics experimentally. The energy transfer to the
Ar part could be measured from the Ar evaporation spectra as long as
one deals with metal clusters embedded in finite Ar drops. But the most
interesting effect is the hindered explosion of the imprisoned Na
cluster and its subsequent shape oscillations in the Ar cavity. Here
we can exploit the pronounced features of the Mie plasmon resonance of the 
embedded metal cluster.
Indeed, it couples strongly to light, the coupling is highly
frequency selective, and the plasmon frequencies are uniquely
related to the cluster shape. This suggests to track the shape
oscillations by pump and probe analysis. In free
clusters, a proper setup allows to map in a unique fashion the
evolution of radial shape \cite{And02} and of quadrupole deformations
\cite{And04}, through the time
evolution of the Mie plasmon resonance. We have computed that for the
case considered here. We take the instantaneous
configuration at a given time and compute the optical response in all
three spatial directions for the actual charge state of the
system. The strong oblate deformation leads to a splitting of the 
resonance peak where the shorter extension along $z$ is associated to
a blue-shift of the mode and the larger extension in orthogonal
direction to a red-shift. 
Our findings are in qualitative agreement with experimental results
obtained for Ag in glass \cite{Sei00}, using a similar pump probe scenario. 
Our model thus provides a microscopic 
interpretation, the first one, to the best of our knowledge, of these 
experimental investigations.

\section{Conclusion}
We have presented a robust model for the description of 
simple Na clusters embedded in an Ar matrix. The dynamics 
of the Na cluster is treated in the TDLDA-MD {\it ab initio}
framework, allowing any dynamical regime (linear and non-linear) 
at the side of the Na cluster. The matrix is treated at 
the simplest level of modeling still accommodating polarization and full 
MD motion at the side of Ar atoms. Both sodium and argon 
are properly coupled following previous 
detailed quantum chemistry calculations. 
On the basis of that model, we 
have investigated the structure and dynamics (low and high energy) 
of the embedded Na cluster. 

We have considered in this paper different systems, namely Na$_8@$Ar$_n$
with $n$ varying from 0 to 434.
The optical response provides an experimental access to the cluster's
structure. When the cluster is embedded, the peak position
results from a 
subtle cancellation between two large effects~: matrix compression of
the ionic configuration, which tends to blue-shift the response, and
electronic polarization effects which tend to red-shift the response.
All in all the net result, in large matrices, comes very close to the
free response. This, by the way, also holds true in the non-linear
domain of the plasmon response where the same cancellation effect is found
to persist. These results are actually quite compatible with related
experiments on Ag in rare gas matrices where it was found that the
plasmon peak is very little modified as a function of matrix size.
The effect of matrix size was also explored in more detail and some results
presented.  Again the effects are relatively small although one
observes specific behaviors for small matrices. Indeed below 164 Ar
atoms we are mostly facing an Ar cluster with an inclusion of a Na
cluster, with properties strongly depending on the system size. For
larger numbers of Ar atoms, we clearly see more regular patterns,
hinting to a convergence of the system's properties towards a
``bulk-like'' limit.

In the last part of the paper, 
we have presented results specifically for the case of non-linear
processes, especially in relation to intense laser irradiation.  The
surrounding Ar stops a Coulomb explosion which would else-wise have
taken place due to the strong laser ionization of the cluster.  
Steady shape oscillations of the cluster are then observed, while the
absorbed momentum spreads 
in a radial sound wave propagating through the atoms and triggers
global radial oscillations thereof. The relaxation
process can be nicely seen in oscillations of cluster radius and
deformation. Both these observables lead to unique
signatures in the plasmon spectrum of the cluster which, in turn, provides
an ideal handle for pump and probe analysis. More 
investigations are in progress to quantify these effects. 

In this paper, we have focused mainly on the actual response of the Na
cluster itself, not discussing deeply the induced impact on the
matrix. It should thus be highly interesting to investigate the impact
of this emission and/or energy deposit on the matrix itself. Our model
was originally tailored for this purpose and it is 
indeed well suited to it. We shall present these results in a
forthcoming publication. Moreover we have
started some research program in the case
of more realistic environments such as MgO or SiO$_2$. Another line of
development, which we also investigate, concerns clusters and molecules
of biological interest, which again have to be treated in relation to
an environment, often water.  Work along that line is also in progress
with aim to consider the effects of irradiation on such systems.

\bibliographystyle{ws-procs975x65}
\bibliography{cmt30}

\end{document}